# On the contrast dependence of crowding

A.Rodriguez,  R.Granger *
*6207 Moore Hall, Dartmouth College, Hanover, NH 03755, United States*

ABSTRACT

Visual clutter affects our ability to see: objects that would be identifiable on their own, may become unrecognizable when presented close together ("crowding") – but the psychophysical characteristics of crowding have resisted simplification. Image properties initially thought to produce crowding have paradoxically yielded unexpected results, e.g., adding flanking objects can ameliorate crowding (Manassi, Sayim et al. 2012, Herzog, Sayim et al. 2015, Pachai, Doerig et al. 2016). The resulting theory revisions have been sufficiently complex and specialized as to make it difficult to discern what principles may underlie the observed phenomena. A generalized formulation of simple visual contrast energy is presented, arising from straightforward analyses of center and surround neurons in the early visual stream. Extant contrast measures, such as RMS contrast, are easily shown to fall out as reduced special cases. The new generalized contrast energy metric surprisingly predicts the principal findings of a broad range of crowding studies. These early crowding phenomena may thus be said to arise predominantly from contrast, or are, at least, severely confounded by contrast effects. (These findings may be distinct from accounts of other, likely downstream, "configural" or "semantic" instances of crowding, suggesting at least two separate forms of crowding that may resist unification.) The new fundamental contrast energy formulation provides a candidate explanatory framework that addresses multiple psychophysical phenomena beyond crowding.  *Keywords:* Crowding, contrast energy, early visual stream

## I.  Crowding phenomena, extant hypotheses, and the nature of visual contrast

### A) On crowding

Perception of an object is strongly compromised by the presence of additional flanking objects within a nearby neighborhood of the target ("crowding"). For instance, identification accuracy of a peripherally-presented letter is severely impaired by the presence of flanking letters; thus crowding is an issue in reading (Legge 2007), as well as many other practical tasks. The crowding effect appears to apply from simple stimuli (lines, letters) to complex objects and complex motion.

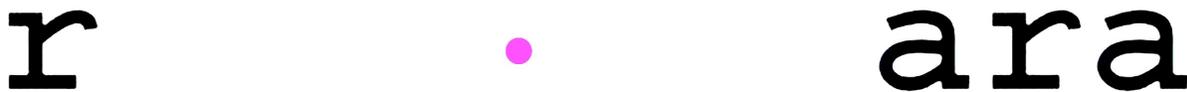

*Figure 1.  Simple illustration of crowding.  Fixating on the central dot, with the page held relatively close to the eyes, the isolated **r** on the left is easy to identify, whereas the crowded **r** on the right is very difficult to identify.*

Crowding is not explained by peripheral reduction of acuity: even when acuity is sufficient to identify an isolated peripheral letter (Figure 1 left), the addition of flankers near the letter impair its identification (Figure 1 right). Indeed, crowding has recently been shown to occur even at the fovea under appropriate experimental conditions (Coates, Levi et al. 2018). Importantly, pure detection (simple presence or absence) of the target remains completely unaffected by flankers; what is impaired is the identification of target features (Strasburger and Rentschler 1996, Pelli, Palomares et al. 2004).

---

Proposed unifying mechanisms have included averaging, substitution, attention, grouping, and more. Many current approaches are based on extensive and detailed simulations of neural elements and architectures, e.g., (Freeman and Simoncelli 2011, Harrison and Bex 2015, Doerig, Bornet et al. 2020). Independent of the validity and importance of the characteristics of these assemblies, it can be unclear how to discern explanatory principles.

Many models entail multiple processing stages that are successively or hierarchically "pooled": early (upstream) neurons with small receptive fields project to those (downstream) with larger RFs, pooling the early information and thus muddling features (e.g., (Freeman and Simoncelli 2011)), in line with computational models (e.g. (Riesenhuber and Poggio 1999)). RF size also increases with eccentricity, enlarging pooling windows in the periphery.

Several findings indicate that seemingly-lost crowded items are not lost entirely; some features remain available to downstream processing, including information about orientation (He, Cavanagh et al. 1996) and texture (Parkes, Lund et al. 2001); these are sometimes interpreted in terms of limited spatial resolution of attention, i.e., a lower limit to attentional window-size at a given eccentricity. Many of these findings lead to suggestions that crowding necessarily arises from cortical operation, rather than from precortical areas along the visual path (Intriligator and Cavanagh 2001, Parkes, Lund et al. 2001, Pelli 2008, Freeman and Simoncelli 2011, Herzog, Sayim et al. 2015). Any proposed early-stage crowding mechanism must be consistent with information that is still shown to be present in subsequent stages (Manassi, Sayim et al. 2013, Manassi and Whitney 2018).

Although added flankers cause the crowding effect, there are seemingly paradoxical cases in which target identification is improved, not impaired, by the addition of more flankers (Wolford and Chambers 1983, Banks and White 1984, Herzog, Sayim et al. 2015) and it has been shown that flankers far outside the usual close neighborhood (Bouma 1970) of the target also can affect crowding (Manassi, Sayim et al. 2012, Manassi, Sayim et al. 2013). It has been proposed (Herzog, Sayim et al. 2015) that these multiple-flanker and distant-flanker effects arise due to "grouping" of flankers together, such that the "grouped" flankers are affecting each other, more than flankers affecting the target. These hypotheses also are consistent with the possibility that some crowding effects may arise from downstream processing far beyond early visual regions. However, a range of possible explanations for these effects remain feasible.

***B) On contrast***
An oft-cited view is that crowding arises from "critical spacing, independent of spatial frequency" (Pelli 2008) and specifically that "contrast" does not suffice as an explanatory mechanism (Flom, Weymouth et al. 1963, Chung, Levi et al. 2001, Levi, Hariharan et al. 2002); (and see (Strasburger, Rentschler et al. 2011) for a partial review).

Multiple contrast measures exist in the literature (such as Weber; Michelson; and RMS contrast (Peli 1990)); correspondingly, "contrast energy" is typically defined as the integral of the square of the contrast over all dimensions in which it varies; see, e.g., (Watson, Barlow et al. 1983, Kukkonen, Rovamo et al. 1993).

Most such measures treat pixels as independent of each other. We proffer a new, "radially generalized" account of contrast energy (of which RMS contrast and others are shown to be special cases). The new account, rather than evaluating contrast pixel-by-pixel in an image, instead formally evaluates radial regions corresponding to receptive fields, within which pixels may have interacting (rather than independent) effects, as viewed by a perceiver.

(This new contrast measure arises from work unrelated to crowding. Studies of the visual dissimilarity between two similar images (such as an image and a degraded or compressed version of the image) led to derivation and analysis of the primary new contrast measure that is also introduced in the present manuscript in Eq.4 below (Bowen et al., 2020).)

The generalized contrast measure is shown to be specifically predictive of the essential results of several well-studied crowding effects from the literature. To reproduce those published results, the sole two steps are 1) measuring contrast energy and 2) mapping it to behavior (the subject's identification of the flanked target). The sole parameter simply maps contrast quantities directly onto behavioral performance, by estimating the threshold at which the contrast has changed enough to begin generating identification errors.

The resulting straightforward measures surprisingly account for multiple instances of crowding across the published literature, including some exemplars that have thus far been resistant to simplification.

At minimum, this is evidence that many standard crowding effects are severely confounded by variations in the introduced contrast measure. We specifically propose that a substantial number of results attributed to crowding, actually arise directly from contrast.

We also provide examples of crowding that are not predicted by contrast. Since there are many clear instances where crowding is predicted by contrast, and instances where it is not, we suggest that attempts to unify crowding to a single phenomenon would presumably have to account for these instances in which contrast alone is explanatory.

Contrast–dependent crowding effects may arise extremely early in the visual stream. Other evidence indicates that certain other crowding effects may arise from later processing; our findings suggest that experiments may be profitably divided into at least two possibly distinct categories: those that are explained by contrast and those that are not. This may indicate that crowding phenomena are not all due to a unitary mechanism. The findings may also help determine which apparent crowding effects are pre-cortical versus cortically dependent. (In addition, the findings may aid in separating attributes of model neural architectures into characteristics that are needed for a particular effect (such as crowding) versus additional neural features that may not be required to explain these phenomena.)

In the following sections, we first introduce the new formal generalized framework for visual contrast; we then use that framework to reproduce the results of a range experiments in the primary literature: (Flom, Weymouth et al. 1963, Pelli and Tillman 2008, Freeman and Simoncelli 2011, Harrison and Bex 2015, Herzog, Sayim et al. 2015).

## II. The differential geometry of visual contrast
### A) Foregrounds, backgrounds, and differences: from early anatomy to early perception

Figure 2a and 2b depict two pairs of three-pixel images (not to scale). The Euclidean distance from image $\vec{a}$ to $\vec{a}'$ is the same as from $\vec{b}$ to $\vec{b}'$ and yet perceivers report significantly more contrast difference between $\vec{b}$ and $\vec{b}'$ than between $\vec{a}$ and $\vec{a}'$, apparently because the slightly contrasting gray center is overwhelmed by the flanking contrast in the $\vec{a}$ pair but not in the $\vec{b}$ pair.

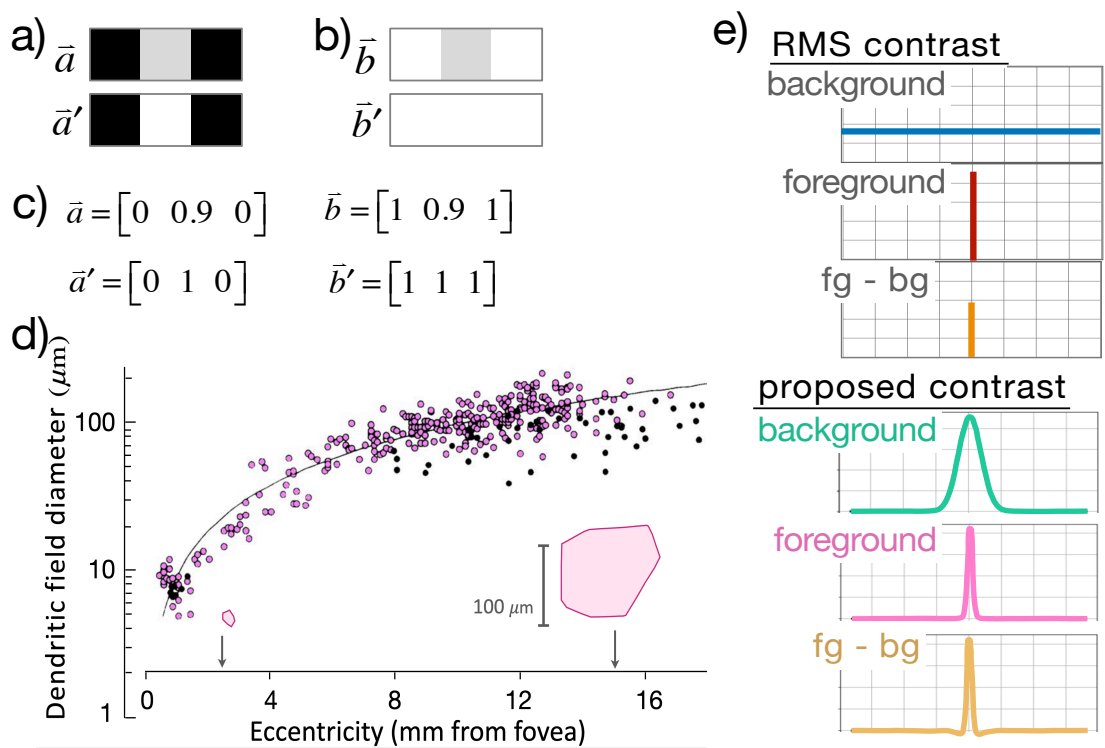

*Figure 2. Different formulations of contrast. **a,b)** Closeup of three adjacent pixels in an image. The intensity difference between center pixels is the same for (a) and (b), but the center pixel is flanked either by dark (a) or light (b) neighboring pixels. **c)** The Euclidean distance between vectors $\vec{a}$ and $\vec{a}\,'$ is identical to the distance between vectors $\vec{b}$ and $\vec{b}\,'$, yet perceivers universally report that the contrast difference between images $\vec{b}$ and $\vec{b}\,'$ is significantly greater than between images $\vec{a}$ and $\vec{a}\,'$. Standard root-mean-squared contrast measures do not account for these perceived differences. **d)** Retinal midget ganglion cell dendritic field diameters increase with their distance from the fovea (logarithmic fit: $y = 8.64 \times x^{1.04} (R = .94)$). Insets are tracings around midget cell dendritic trees at 2.5 and 15 mm eccentricity. (Scale bar = 100 μm.) (Figure adapted from (Dacey and Petersen 1992)). **e)** Depictions of standard RMS contrast (top), versus contrast as determined by receptive field center and surround at a given eccentricity (bottom). The "foreground" of RMS contrast is a single selected pixel, and the background is the entire visual field; the foreground and background determined by cell dendritic field sizes are approximated as Gaussians, with contrast determined by the difference of these Gaussians (*fg-bg*).*

Standard measures of visual contrast, such as root-mean-square (RMS) contrast (Peli 1990), treat pixels independently, and thus yield equal contrast differences for the paired images in Figure 2a and 2b, as per the Euclidean vector renderings of the images (2c). These contrast difference measures clearly disagree with human judgments of the perceived difference between these two contrasts, which appear influenced by the context of surrounding pixels.

The different human perceptual evaluations in Figures 2a and 2b may readily arise from cell properties in the early visual pathway from the eye to the brain; beginning in the retina, and continuing through subsequent visual processing stages, neurons exhibit center-surround receptive fields, which are classically modeled as the difference of two overlapping Gaussian profiles with comparatively larger and smaller diameters. (Although this discussion will focus on retinal neurons, similar arrangements of center and surround cells occur throughout the early visual stream.) (Rodieck 1965, Young 1987, Dacey and Petersen 1992, Wandell 1995, Dacey 2004).

Figure 2(d) shows the dendritic field diameters of retinal midget cells near and far from the fovea. Empirically, psychophysical studies have referred to an approximate fixed ratio ($k=5$) of center to surround across a range of eccentricities (Young 1987).

Differences of Gaussians (Figure 3c) form bandpass filters (Figure 3d). If these are simply applied to images such as Figure 3b, the results are seen in Figure 3e. This is the processing that is proposed to occur if the image ("ara") in Figure 3b appeared in the periphery at a specified visual angle of eccentricity away from a fixation point, as in Figure 1.

The bottom row of Figure 3(c,d,e) describes the overall operation that defines the new proposed contrast operation that arises from this straightforward retinal processing. It is notable that the resulting contrast sensitivity profile (Figure 3d, bottom) corresponds to measured human spatial frequency sensitivity (Wandell 1995) (Figure 3a).

(Qualitatively, it is notable that the gain is non-monotonic with respect to spatial frequencies in images, both in the center-surround model (Figure 3d) and in empirical measures in humans (3a). Specifically, there is a spatial frequency "sweet spot": portions of images whose features have a spatial frequency in roughly the 3-10 Hz range are perceived with (slightly) higher acuity than features with higher or lower spatial frequencies. It will be seen that this, too, is consistent with data from crowding experiments (Flom, Weymouth et al. 1963), in section III of this article.)

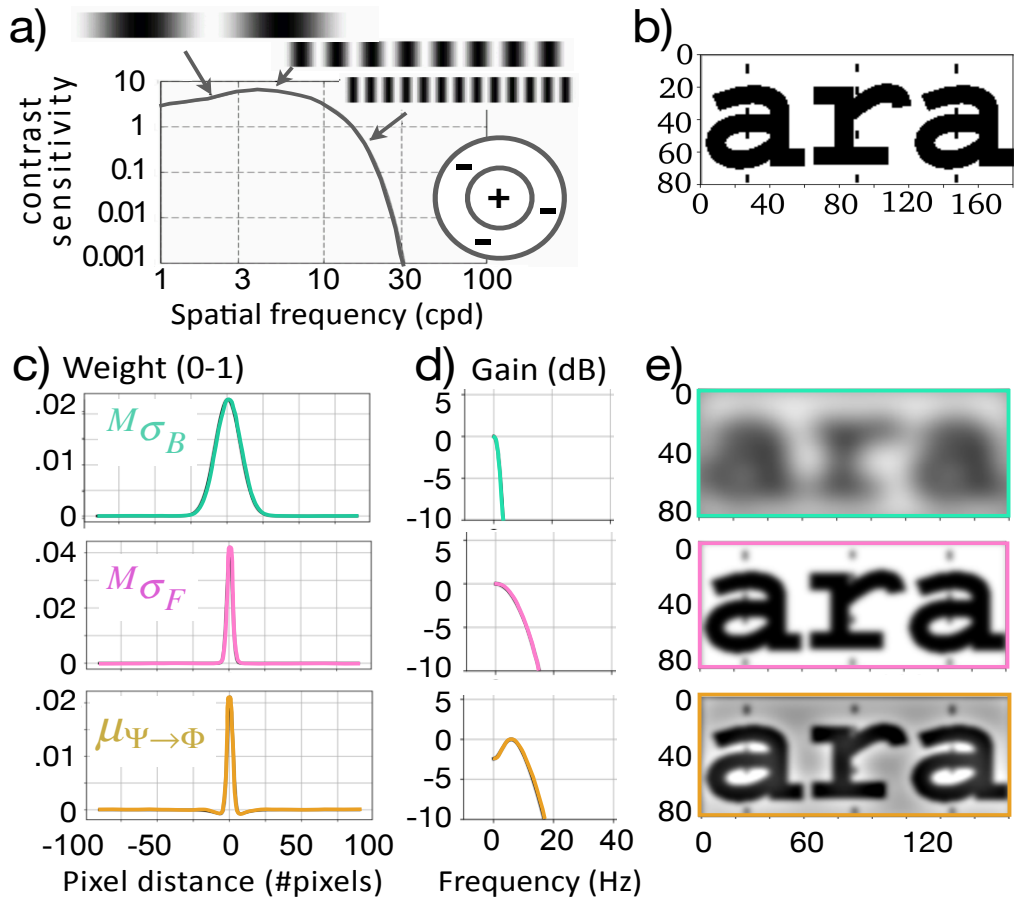

Figure 3. (a) Sample measures of contrast sensitivity as a function of visual spatial frequency in human subjects (Wandell 1995). (b) Original image presented in the periphery at a fixed visual angle distance from fixation point as in Figure 1. (c) Calculated background and foreground weight masks (broad and narrow Gaussians, respectively), derived from retinal cell properties (Figure 2d) and their difference of Gaussians (Equation 3 below). (d) Differential gain by frequency band, arising from the specified weight masks; background and foreground alone describe monotonic frequency gains, yielding restrictive and less-restrictive (background, foreground) low-pass filters; the difference of Gaussians (bottom; yellow) is non-monotonic, describing a bandpass filter, as in (a) above. (e) Results of these operations applied to the original image in 3(b) and in Figure 1.

### B) Formal derivation of contrast from biological cell properties

From the biological features in the previous section, we formalize a measure of visual contrast that incorporates radially adjacent pixel zones, due to the center-and-surround characteristics of cells in given visual receptive fields. The resulting treatment arrives at *radially-generalized* contrast metrics that are in agreement with simple human judgments of this kind as in Figure 2a and 2b.

We first designate specified radial regions around any given pixel, as per the cell dendritic radii of Figure 2d, here termed "foreground" and "background," as in Figures 2e and 3c.

We define a foreground "mask" $M_{\sigma_F}$ constituting a set of weights assigned to each pixel within the region. We assume that these weight functions take the form of a Gaussian with a small sigma $\sigma_F$ as follows:

$$M_{\sigma_F} = (2\pi\sigma_F^2)^{-1/2} e^{-\left(((p-x)^2 + (q-y)^2)/2\sigma_F^2\right)} \tag{Eq.1}$$

where $p$ and $q$ are the indices for the values of the superimposed Gaussian receptive field that is centered at pixel location $x, y$ with standard deviation $\sigma_F$. (We note that these fields need not take Gaussian form; other structures will be briefly discussed, but we use these for the results in the present work).

The background mask function has the same form, but with larger standard deviation $\sigma_B$ which is defined as a constant multiplied by $\sigma_F$ ($K = \sigma_B / \sigma_F$),

$$M_{\sigma_B} = (2\pi\sigma_B^2)^{-1/2} e^{-\left(((p-x)^2 + (q-y)^2)/2\sigma_B^2\right)}$$

and we combine these to define a difference-of-Gaussians convolution mapping:

$$\varsigma(x, y, p, q) = M_{\sigma_F} - M_{\sigma_B} \tag{Eq. 2}$$

This will be convolved with a given image to produce the convolution map that will be used to derive the radially-generalized contrast quantity.

### C) The contrast of an image

For a given resolution, an image consists of the intensities of $d$ pixels; these define a corresponding $d$-dimensional image vector space $II$, such that an image $i$ defines a vector $\vec{v}$ in $II$: $\vec{v} = vec(i)$ where the $i$ values are intensities at each pixel. Any such vectorization treats pixels as independent, losing their neighbor relations in the 2d image.

To account for neighbor relations in space $II$, we define a Jacobian $J$ (a $dxd$ matrix) that performs a convolution, such that convolving the difference-of-Gaussians operator from Eq.2 with the image $i$ ($\varsigma * i$), and then vectorizing the result ($vec(\varsigma * i)$), is equivalent to first vectorizing the image and then multiplying by the Jacobian; that is:

$$J \cdot vec(i) = vec(\varsigma * i) \tag{Eq. 3}$$

The Jacobian thus incorporates radial regions in an image vector $\vec{v}$. We define the "Radially Generalized contrast energy" semi-norm, $C(\vec{v})$, as follows:

$$C(\vec{v}) = (\vec{v}^T J^T J \vec{v}) \tag{Eq. 4}$$

(This satisfies the semi-norm properties: it is positive $\forall \vec{v} \in II$; it is absolutely homogeneous, i.e., $C(\alpha\vec{v}) = |\alpha| C(\vec{v})$; it is subadditive, i.e., $C(\vec{u} + \vec{v}) \leq C(\vec{u}) + C(\vec{v})$. Thus it induces a semi-metric. For purposes of this article, we simply will refer to it as a metric.)

This generalized metric may be compared with standard definitions of contrast energy; see, e.g., (Watson, Barlow et al. 1983, Kukkonen, Rovamo et al. 1993). It is instructive to recognize the special case in which the foreground $M_{\sigma_F}$ is reduced from a Gaussian to have simply a value of 1 at a single pixel, and zero for all other pixels; and the background $M_{\sigma_B}$ is the uniform distribution with values *1/n* at all *n* pixels in the image. (See figure 2e, top). For this degenerate $\varsigma$ mask, the $C(\bar{v})$ operation would produce a simplified degenerate instance, from which other contrast measures fall out directly as special cases; in particular, RMS contrast is the square root of generalized contrast energy per areas, i.e., $RMS = \sqrt{C(\bar{v})/A}$, where **A** can be treated as unitary for the present cases. This is one introductory instance of contrast, to show that RMS-contrast falls out as a special case from the broadened formulation of the contrast energy semi-norm presented here.

### III. Five representative crowding experiments
#### *A) From contrast to behavior*
In the following published crowding experiments, subjects were presented with an image in which each object is designated by the experimenter as either target or distractor, and subjects were instructed to identify the target, with or without the presence of distractors.

We test the hypothesis that, for these experiments, high values of the contrast energy metric predict improved target identification performance in the presence of the distractor, whereas low contrast energy predicts worse target identification performance. The latter is what is referred to in the experiments as a crowding effect.

In general, formal expressions that predict behavior contain at least one experimental parameter that corresponds to the mapping of a psychophysical calculation to explicit behavioral measures. (Many such experiments contain multiple parameters, each of which is in some way fitted to observed data.)

In the present model we wish to identify as closely as possible those response accuracies that can be said to arise directly from the contrast metric; thus we wish to map the (internal) contrast calculation to the (overt) behavioral measure of correct target identification, via as few parameters as possible. It should be noted that this will have the effect of a) directly implicating the contrast metric in the observed behavior (while avoiding overfitting), as well as b) possibly demonstrating some behaviors that are not predicted by contrast, but presumably by other psychophysical operations outside the scope of the present study. We will show instances of both behaviors that are, and behaviors that are not, accounted for by contrast, with the aim of separately characterizing these distinct psychophysical behaviors.

We thus introduce a single parameter, $E_\alpha$, the contrast energy value at which a subject's ability to identify the target among distractors begins to become impeded by the flanking distractors in an image. This is the point at which a given experiment exhibits the subject's sensitivity to the effects of the distractors on the target.

We construct a gaussian that maps contrast to proportion of correct subject response to the target alone, and a gaussian for when the target is presented with flanking distractors.

In isolation, a target has a given measurable contrast value $\mu_\tau$. We assume that a 10% change to that target contrast value will yield a recognition error rate of 0.01, i.e., 1% misidentifications of the altered target. (More empirically-fitted figures would normally be arrived at from experimental findings. As emphasized below, we do not consider any features or configurations of images whatsoever, nor are we attempting to match exact performance. We simply assume a tight range of contrast around which the target is identifiable, and that identification errors begin to arise at 10% contrast change. These straightforward simplifying assumptions are highlighted in order to spotlight the surprising ability of these simple contrast metrics to predict certain crowding findings.)

These values are used to calculate a "target-alone" Gaussian distribution $\mathbb{G}_\tau$ with a mean of $\mu_\tau$ and a s.d. $\sigma_\tau$ (defined below), such that the distribution's value drops to 1% at a point that is at either 0.9 $\mu_\tau$ or 1.1 $\mu_\tau$, i.e., 10% from $\mu_\tau$. (Figure 4; green Gaussian). (These assumptions for $\mathbb{G}_\tau$ are fixed parameters, that do not depend on the experiment or on any 'curve-fitting'.)

$$\mathbb{G}_\tau(x) = e^{-(x-\mu_\tau)^2/2\sigma_\tau^2}.$$

For x = 0.9 $\mu_\tau$ or for x = 1.1 $\mu_\tau$, then $\mathbb{G}_\tau(1.1\mu_\tau) = e^{-(1.1\mu_\tau - \mu_\tau)^2/2\sigma_\tau^2}$.

For the conditions when the target is presented with flanking distractors, a "target-plus-flankers" Gaussian distribution is calculated, $\mathbb{G}_\phi$ (see Fig 4a), that shares the mean $\mu_\tau$, and has a s.d. $\sigma_\phi$ such that the Gaussian's value drops to 1% at whatever point in the experiment that a flanker elicits no crowding; i.e., we assume that for that particular stimulus image, that the target will be identified 99 percent of the time. (The value $k_\phi$ is defined below).

$$\mathbb{G}_\phi(x) = e^{-(x-\mu_\tau)^2/2k_\phi^2\sigma_\tau^2}$$

That point, again, constitutes the single parameter in the model that arises from the experiment itself: the contrast value $E_\alpha$, at which a subject's ability to identify the target first begins to be impeded by distractors (see Figure 4a).

$$\mathbb{G}_\phi(E_\alpha) = e^{-(E_\alpha - \mu_\tau)^2/2k_\phi^2\sigma_\tau^2}$$

To compute the value of $\sigma_\tau$, the standard deviation of the Gaussian $\mathbb{G}_\tau$, we recall the assumption that a 10% change to the target will correspond to 1% successful identification of the target, and solve for $\sigma_\tau$:

$$\mathbb{G}_\tau(1.1\mu_\tau) = e^{-(1.1\mu_\tau - \mu_\tau)^2/2\sigma_\tau^2}$$

$$0.01 = e^{-(1.1\mu_\tau - \mu_\tau)^2/2\sigma_\tau^2}$$

$$\ln(0.01) = -(1.1\mu_\tau - \mu_\tau)^2/2\sigma_\tau^2$$

$$\sigma_\tau^2 = -(1.1\mu_\tau - \mu_\tau)^2/2\ln(0.01)$$

We define a constant $h=2\ln(0.01)$ (and note that h<0); then

$$\sigma_\tau = \left[-(1.1\mu_\tau - \mu_\tau)^2/h\right]^{1/2}$$

For $\sigma_\phi$, similarly,

$$\mathbb{G}_\phi(E_\alpha) = e^{-(E_\alpha - \mu_\tau)^2/2\sigma_\phi^2}$$

$$0.01 = e^{-(E_\alpha - \mu_\tau)^2/2\sigma_\phi^2}$$

$$\ln(0.01) = -(E_\alpha - \mu_\tau)^2/2\sigma_\phi^2$$

$$\sigma_\phi^2 = -(E_\alpha - \mu_\tau)^2/h$$

$$\sigma_\phi = \left[-(E_\alpha - \mu_\tau)^2/h\right]^{1/2}$$

(Having computed $\sigma_\tau$ and $\sigma_\phi$, we can express $\sigma_\phi$ in terms of multiples of $\sigma_\tau$ by defining $k_\phi$ such that $k_\phi = \sigma_\phi/\sigma_\tau$. (The values used for each of these parameters, for each of the experiments analyzed, appear in Table 1, near the end of the article.))

In total, the predicted proportion correct (target-alone and target plus flanker trials) is calculated from examination of the table in Figure 4b:

$$p = \mathbb{G}_\tau + (1 - \mathbb{G}_\phi) \qquad \text{(Eq. 5)}$$

(It is worth noting that these Gaussians used for purposes of estimating subject response, are not related to the Gaussians that comprise the center-surround organization in the early visual pathway that are used to derive the generalized contrast metric.)

In sum, low values of this formula predict a low percentage of target identifications by the subject; high values predict a higher proportion of identification success. (In the experiments reported here, all values are scaled from percentage of 0 to a maximum percentage correct of 85%, i.e., all results *p* are multiplied by 0.85. This arises solely from the data that appears throughout the cited experiments from the literature; those experiments are apparently calibrated such that subjects tend never to get 100% correct recognition rates, but, rather, their ceiling occurs at roughly 85% empirically. These may arise from other factors such as resolution, distance, brightness, all of which contribute to the ultimate ability of the subject to perform the task correctly; the experimenters may have calibrated the tasks so as to avoid ceiling effects. Empirically, in these experiments that 85% ceiling is thus not related to crowding effects per se, but rather is the best that the subjects can do when the targets are not crowded. Our model thus is simply calibrated to that empirical ceiling from the literature (as with typical parameters of other models), in order to compare model results to subject results.)

We provide this somewhat extensive derivation of Eq 5 to make it clear that the only experiment-derived parameter that appears anywhere in the calculations is that of $E_\alpha$; and that Eq 5 is then derived according to usual principles of classification using a contingency table as in Figure 4b. Eq.5 then straightforwardly maps the contrast energy of an image to the proportion of a subject's correct responses.

Note that the full code for computing all of the calculations in this article, and reproducing all of the material for the figures, is available on github:
https://github.com/DartmouthGrangerLab/Contrast/

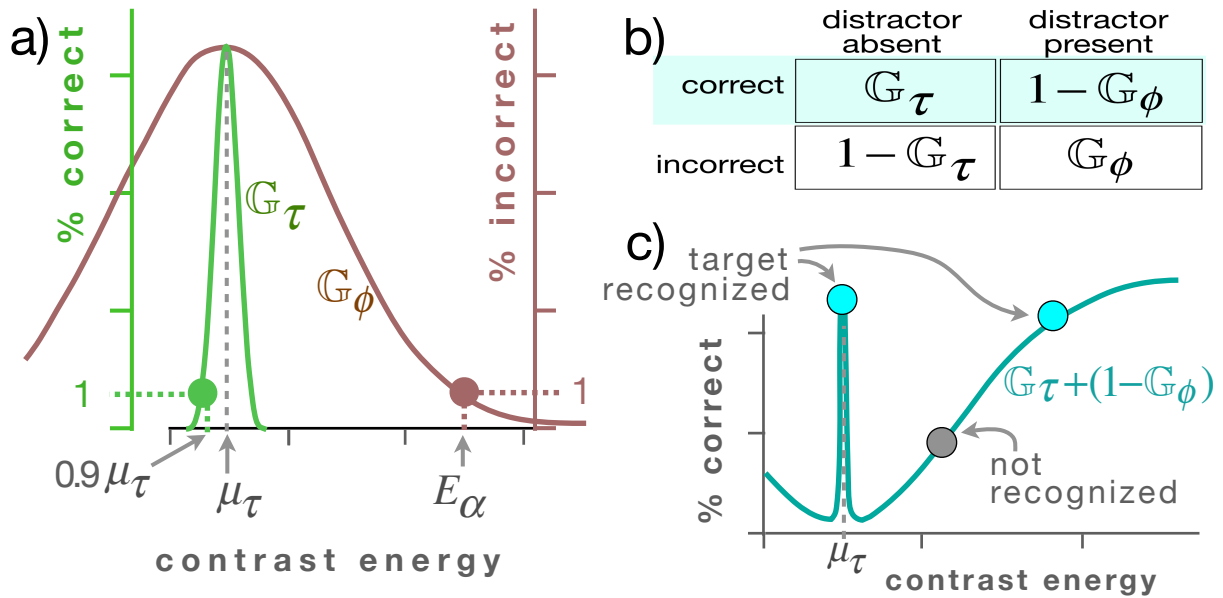

*Figure 4.  a) A gaussian $\mathbb{G}_\tau$ that maps contrast energy (from Eq 4; x axis) to proportion of correct subject response to the target alone (green y axis), and a gaussian $\mathbb{G}_\phi$ that maps contrast to proportion of incorrect subject responses to the target in the presence of distractors (brown y-axis). Also shown is the parameter $E_\alpha$, the contrast energy value at which the target is identified 99% of the time by a subject in a given experiment.  b) The four types of subject responses: correct identification of target alone; correct identification of target in the presence of flanking distractors; failure to identify target; failure to identify target in the presence of distractors. Highlighted are the two forms of correct identification; their sum comprises the overall estimate of correct identification trials.  c) The function that is the sum of the two types of correct identification trials from (b), i.e., $p=\mathbb{G}_\tau+(1-\mathbb{G}_\phi)$; high values predict higher percentage of correct identifications (blue points); lower values (gray) predict smaller percentage of correct identification.*

The relevance of these estimations of subjects' response accuracies can be intuitively understood by noting that recognizing is not simply perceiving: it is matching the perceived entity against some stored version.  That is the difference, for instance, between perceiving that there are pixels present, versus recognizing that they take a form that has previously been seen by the subject.  Every crowding experiment presented here (possibly all such experiments in general) rests on the presupposition that the subject can identify whether or not the seen target entity is the "same" as some previously seen entity, whether that was long-ago acquired (e.g., a typed alphabetic letter) or indicated to the subject in the instructions (e.g., the target Landolt C's angle, versus flanking Landolt C's angles).

Within a given task, a specific target is associated with a specific value of its measured contrast energy.  If the target were presented alone, then potential targets that deviate from the task's target value would be harder to identify, impairing performance, whether the contrast value is increased or decreased.  (Matching a target alone does not arise in the experiments modeled here).  When target plus flankers are presented, the subject must identify which pixels in the image represent the instructed "target", and report on its characteristics (e.g., its name "r" or its gap angle "90 degrees").  The more distant the overall contrast energy is from the target-alone contrast energy, the easier it is for the subject to distinguish the target within the interfering flankers.  The closer the flankers are to the target, the closer the contrast energy of the overall image is, causing reduced predicted correct response rates.

For a given image from each of the experiments in the following section, the eccentricities of the pixels in the relevant peripheral region were calculated given the reported details of the experimental setup.  The retinal coordinates for the corresponding image pixels were computed using the viewing distance and screen resolution used in the specified experiment.  Midget cell diameters (Dacey and Petersen 1992) for those retinal coordinates were collected with a 17.2° width around the region of interest.

We again emphasize that the present work entails no analyses of any detailed feature or configuration characteristics of any kind, such as shapes or orientation.  The sole quantity tested is the newly introduced contrast metric.  Thus no specific experimental results from the crowding literature are addressed with respect to description of a target object, its orientation, or other features.  Rather, each treatment of an experiment simply proceeds by first computing the contrast energy for the images used (following the methods from the previous sections), and then using the mapping of Eq 5 (Fig 4) to calculate the efficacy with which the computed contrast energy can determine which visual entities constitute the target and which do not.

This procedure would thus seem to be utterly insufficient to capture crowding findings, which, after all, appear to entail subjects' identifying detailed features of an image (such as the orientation angle of a Landolt C).  It is thus illuminating to show that contrast accurately predicts the accuracies of subjects' responses, despite the fact that contrast has no information whatsoever about orientation or other configural attributes of the image.  This suggests at minimum a substantial role for contrast in these reported findings in the literature.

For each experiment below, both target-alone and target-plus flanker images are viewed, and contrast is computed for each.  The figures show a) experimental stimuli; b) calculation of contrast energy from flanker distance; c) calculation of predicted identification accuracy from contrast energy; and d) combining b and c, calculation of predicted information accuracy from flanker distance, which is then compared alongside corresponding measures from the reported literature.

Again: lacking any features, orientations, or other attributes of the images, the model produces no specific image characteristics; rather, it determines the calculated contrast energy of the image under the given viewing conditions and determines, solely from this value, whether the image will be recognized.

(It is further noted that, in addition to the direct predictability of the experimental results from the new contrast metric, several of the experiments are predictable even with the simpler standard RMS contrast quantity. For instance, the experimental materials in Pachai et al. create flanking Landolt Cs that surround a central C; as those surrounds grow larger, they add more pixels to the image, which increases generalized contrast, and also increases simple RMS contrast in the image. (Table 2, below). Controlling for these factors is required to separate contrast-dependent from contrast-independent crowding effects.

The following sections detail the findings of four specific instances of crowding studies. These provide a simple range of basic crowding results illustrating the dependency of the effect on flank distance and eccentricity, across a span of visual images with different features. What is seen is that contrast energy (and thus, contrast) alone is highly predictive of the recognizability of the crowded target objects.

***B) (Pelli and Tillman 2008, Freeman and Simoncelli 2011)***
We first illustrate the essence of the effect with some well-studied findings from several researchers, in which a range of tests illustrate the crowding effect in the context of its two primary parameters: eccentricity of the target object from the fovea, and distance of neighboring flanking objects from the target object. The targets in such experiments range from printed letters to images, placed at a range of distances (eccentricities) from a fixation point, with flanking images at differing distances from the target (Figure 5a). We show results solely from a set of letters (as both target and flankers) at a range of eccentricities and flank distances.

Figure 5b shows calculated "heatmaps" of convolved sample stimulus images (see Eq 3) in particular configurations (all with surrounding "a" images at the fixed flank distance of 0.2° from the central target "r", centered at eccentricities of 5, 10, 15, and 20 degrees of visual angle from the fixation point. As described, the receptive fields at these locations are calculated from the dendritic radius of midget cells at a given eccentricity (Fig 2). The convolved images illustrate how the radial interactions among neighboring pixels are intensified with increasing eccentricity; nearer the fovea, the images exhibit higher generalized contrast, and their edges can be more clearly seen; as they recede from the fovea their edges decrease in clarity.

Figure 5c plots the calculated contrasts of the sets of images for a range of eccentricities from the fovea, and flank distances from the central target.

Intuitively, the higher the eccentricity (distance from the fovea), the larger the standard deviation of the foreground Gaussian, and thus the lower the resulting calculated contrast energy for the overall target-plus-flanker image.

Reciprocally, at a given eccentricity, changing the flank distance has little effect on contrast; the exceptions are those flankers that are comparatively quite close to the target: then contrast becomes reduced, largely because the foreground Gaussian is averaging more content into a smaller region. (Whereas quantities such as RMS contrast do not take the neighborhood of the target into account, the new contrast energy metric does so.)

As described, the parameter $E_\alpha$ is the contrast energy value of the overall image, at which subjects exhibit 1% error in target identification. That parameter is thus used to map the calculated contrast energies to estimations of the proportion of correct identifications that a subject will exhibit. Intuitively, $E_\alpha$ is the high contrast value at which the target-recognition curve (from Figure 4) will begin to dip below its maximum proportion-correct value; contrasts lower than $E_\alpha$ predict fewer correct identifications (Figure 5d). Then graphs (c) and (d) are combined to produce the standard type of reports for such experiments, with flank distance on the x axis and proportion correctly identified on the y axis (Figure 5e).

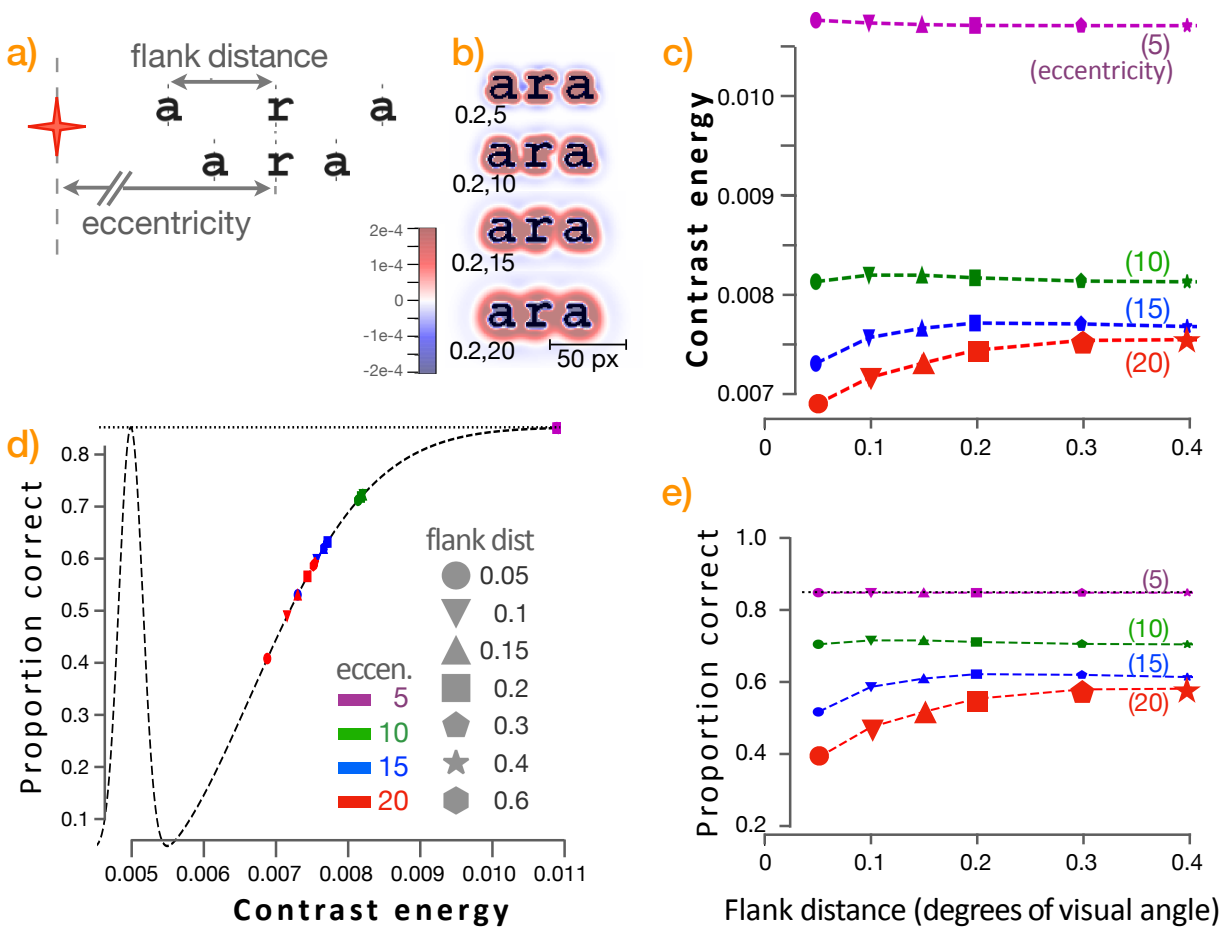

*Figure 5. **a)** Target objects (letters large and small; the letter "r" for this example) at different eccentricities from a fixation point (red), are surrounded with non-target objects at different flanking distances. **b)** Heat maps show several sample images with the same flank differences, but different eccentricities, convolved with the corresponding difference-of-Gaussians Jacobian (Eq 3). The lower the eccentricity, the less radial contamination from neighboring pixels in the convolved image. **c)** Generalized contrast energy (Eq. 4) is computed from flank distance (0.05 to 0.4° of visual angle) at several different eccentricities (5°-20° of visual angle). Even in the presence of flanking objects, contrast is relatively stable until those flankers encroach on the target (closer than ~ 0.15° from the target. Less flanker-induced contrast reduction occurs at eccentricities closer to the fovea. **d)** From the flanker-dependent and eccentricity-dependent contrast values, the mapping function (Eq. 5) calculates estimated proportion of correctly-identified targets. At lower contrast energy values, the effect of different contrasts on correct identification is steeper, i.e., slight changes to contrast energy can substantially change estimated target identification rates. At higher contrast energy values, the effect saturates. **e)** Combining c) and d), the predicted ability to identify a target is shown as a function of flank distance and eccentricity.*

### C) (Flom, Weymouth et al. 1963)

A similar analysis was performed on the results of Flom et al (1963), whose target was an annulus with a gap (a Landolt C) at various rotations. In these experiments, the task was not just to identify the target, but to measure the angle at which the gap appeared (i.e., a direction on the surface of the C). The flankers were bars at various distances (Figure 6a). The images were shown at substantial distances from the viewers, with flankers at correspondingly distant locations; these distances are measured not in degrees of visual angle, but in minutes of visual angle. Figure 6e shows the new contrast analysis alongside some of the original findings (inset). It can be seen that the contrast energy metric qualitatively (and somewhat quantitatively) corresponds to the empirical findings in human subjects – despite the fact that humans are reporting on rotation angle, whereas there is no angle information whatsoever contained in the contrast metric.

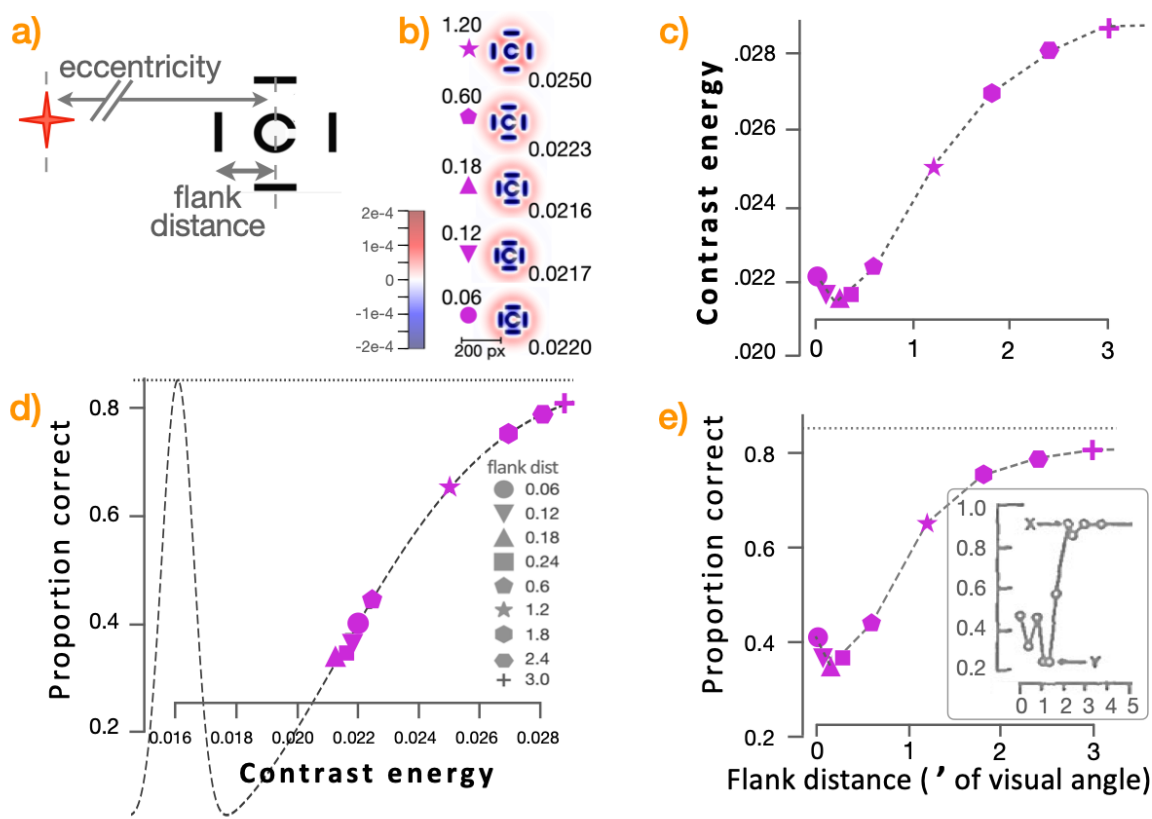

*Figure 6.  a) Flanking bars at different distances from a target Landolt C (all at fixed eccentricity of 3° of visual angle from fixation point); the task was to identify the angle of rotation of the C.  b) Heat maps of selected images with flanks at particular distances (left side; corresponding to x axis of c) and e), exhibiting particular contrast energies (right side, corresponding to y axis of graph c)).  At flank distance 1.2', separations can be seen (red color) between flanks and target, whereas closer flanks interact with the target.  c) The contrast energy semi-norm (Eq.4) is computed from a range of flank distances (at fixed eccentricity).  At extremely small flank distance (0.06', measured in minutes of visual angle, i.e., units of 1/60 of a degree of visual angle), the computed*

*contrast is relatively low (roughly 0.220).  As flank distances slightly increase, contrast energy decreases (to a minimum of roughly .0216).  Beyond this "sweet spot" (see Figure 3a), contrast energy monotonically increases as flankers recede from the target object, reaching an asymptotic contrast value of about 0.29 once flank objects are about 3' distant from the target object.  d) From the flank-distance-dependent contrast values, the mapping*

*function (Eq.5) calculates the estimated proportion of correct identifications.  At lower contrast values, the effects of contrast differences are steep, i.e., slight changes to contrast can substantially change estimated target identification; at higher contrasts, there is little effect.  The subtly non-monotonic nature of the contrast effect is*

*seen: flank distance of 0.06' yields roughly a 41% estimated correct identification rate, but further increasing the flank distance (to .12, .18, and then .24) cause reduction in contrast, and in identification rates. Increasing flank distance further still (to 0.6' and above) increases contrast and correspondingly increases rate of correct object identification (see Figure 3a and 3d). **e)** Combining c and d, the ability to identify a target is shown as a function of flank distance (arising solely as a function of the effects of flank distance on contrast, as in part c). (**Inset** in e) Results from original experiment of Flom et al. (1963); shown are flank distances in minutes of visual angle and proportion of correct answers. The mark x denotes the maximum flanker separation at which effects of flankers affecting identification (i.e., crowding) are seen; y denotes the flank distance producing maximum crowding (Flom, Weymouth et al. 1963).*

(Note that in (6b), the figures exhibit slightly different colors, due to the fact that the Landolt C and flanker images were black against gray (not shown in 6a), as opposed to gray on white for the previous experiment as in Figure 5a.)

Again, it is noteworthy that in these contrast and mapping calculations, there are no image features such as the Landolt C gap, nor its angle. The mapping function merely uses the variance in contrast energy of the image, to calculate the estimated proportion of correct identifications that the subject will achieve, as described in Figure 4.

Thus the experimental analyses in these sections show at least some direct confounds with the newly introduced general contrast energy metric, or, put differently, the findings indicate that certain crowding effects may be arising predominantly from radially generalized contrast energy.

Moreover, as will be seen, some experiments are confounded not solely with the new contrast metric, but also with standard RMS contrast. The previous experiments (Pelli & Tillman; Flom et al.) are not confounded with RMS contrast. Nonetheless, the radially generalized contrast metric introduced here still predicts the results. These apparent confounds are listed in Table 2.

### *D) (Manassi, Sayim et al. 2012)*

From multiple reported findings from these researchers, we focus on a surprising subset in which the target is a pair of slightly unaligned bars (a vernier), flanked by varying numbers of bars (Figure 8), which resulted in a seemingly paradoxical finding: although adding flankers around a target object did cause crowding, i.e., impaired recognition of the target object (the vernier), the further addition of more flankers unexpectedly ameliorated that effect. An increased number of flanking bars paradoxically caused less crowding than did fewer flankers (Manassi, Sayim et al. 2012).

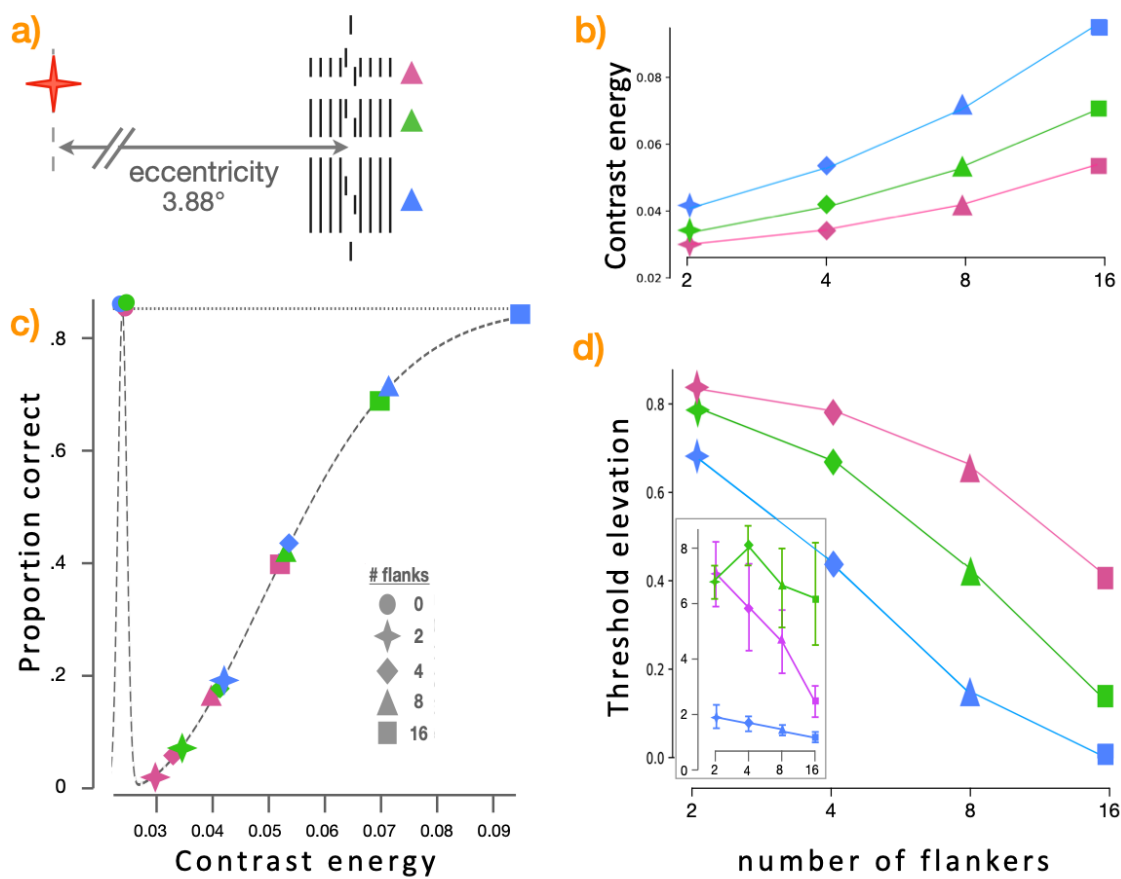

*Figure 7.* ***a)*** *Varied numbers of flanking lines around a fixed target vernier whose two components were either left-top, right-bottom (shown), or right-top left-bottom; that distinction is what human subjects were asked to report. Experimental configurations replicated herein were either two, four, eight, or sixteen lines, surrounding the target vernier (all at a fixed eccentricity of 3.88 visual degrees from the fixation point).* ***b)*** *As before, the calculated contrast energy metric (Eq.4) is shown for each flank configuration, as in the legend (shown in inset table in part (c)). Contrast energy is higher for larger flankers (blue>green>red), and increases with added flankers (2,4,8,16).* ***c),d)*** *This relationship continues to obtain across the range of mapping calculations (Eq.5) generating predicted recognition success rates given numbers of flankers of various sizes. Low "threshold elevation" predicts little to no crowding; larger threshold elevations indicate crowding, i.e., degradation of recognition performance. The inset in d) shows corresponding reported findings from Manassi et al. (2012), some of which may disagree with the predictions from the contrast metric; these are further discussed below.*

Multiple issues arise in the findings reported here. First and foremost, a set of experiments from Manassi et al. that are not shown here entail a set of flanking lines that are "jittered" – instead of aligned at top and bottom as in all other conditions, these "jittered" lines start and stop at different vertical heights throughout the "flanking" region: their appearance is far more like additional verniers, than like flanking lines. The resulting reported threshold elevation results are also wildly unlike those of all other experimental conditions: rather than residing in a range of roughly 2 to 8,

"jittered" flanks rise to values of approximately 30: roughly four times the effect of other conditions. As Manassi et al. suggest, this is evidence of further perceptual processes beyond standard approaches. The effects of these "jittered" stimuli appears to be clearly outside the scope of predictions from the contrast energy metric.

Even in the stimuli reported here, there are potential differences between the contrast-based predictions and the empirical findings from Manassi et al. (Fig 7d plus inset). The contrast metric predicts that adding flankers that are shorter (red), the same height (green), or taller (blue) than the vernier will all ameliorate crowding (descending lines in Fig 7d), whereas Manassi et al. report concordant effects for shorter or longer flanks, but not for venier-sized flanks (7d inset); it is notable that addition of their vernier-sized flanks does trend somewhat toward crowding reduction in their results, though the size of the error bars make it possible that these results may or may not agree with the predictions presented here; overall, our ability to specifically predict their findings is equivocal, and suggests the potential value of further studies on such images.

### *E) (Harrison & Bex 2015; Pachai, Doerig et al., 2016; Harrison & Bex 2016)*

The target image is an oriented Landolt C, as in (Flom, Weymouth et al. 1963), now with concentric surrounding flanker Landolt Cs, of differing radii, differing gap orientations, and sometimes no gaps. The task is to report the angle of the target C, in the face of sometimes-conflicting angles from flanking C's.

As emphasized, the contrast metric computed here yields no information whatsoever about most visual features, such as shapes, gaps, angles. The computations solely indicate the value of contrast energy (as calculated via Eq.4), independent of feature configurations. Yet as has been seen in previous examples, those contrast calculations repeatedly generate predictions of recognizability of images, despite doing so in the absence of the features of the images themselves, and we so far have seen several cases in which those predictions match empirical findings (Figures 5,6,7). This again suggests the possibility that extremely simple visual characteristics may be responsible for differences in reported image recognition errors in those experiments.

The reported empirical results in Figure 8e, below, shows that a flanker containing its own gap causes more interference with correct target angle recognition than a flanker with no gap (a concentric circle). More intriguingly, the researchers found, in concord with the findings of Manassi et al. (2012) above, that whereas recognition of a target is impeded by a flanking object, that interference paradoxically is lessened, not increased, by adding further flankers.

Note that these flankers have a characteristic not present in the flankers from other experiments: when an encircling flanker is moved further from the target, the flanker gets larger and thus contains more pixels; that increases the contrast energy of the overall image, and increased contrast energy predicts improved target recognition in the presence of crowding flankers; this potential confound may affect subjects' success rates.

As with the previous analyses, the angle of the target is not calculated or reported here. What is predicted by the contrast calculation is the rate of failure of subjects to correctly report this feature (gap angle) of the target (or, possibly, any specific feature present in the target). (As the authors of the studies show, empirically the subjects either recognized the target C and responded correctly, or they failed to distinguish the target C, in which case they reliably reported instead the angle of the confounding flanker. These findings are consistent with the mapping process reported here (Eq 5), in which the formula distinguishes between contrast characteristics of the target alone, versus contrast characteristics of other image constituents.)

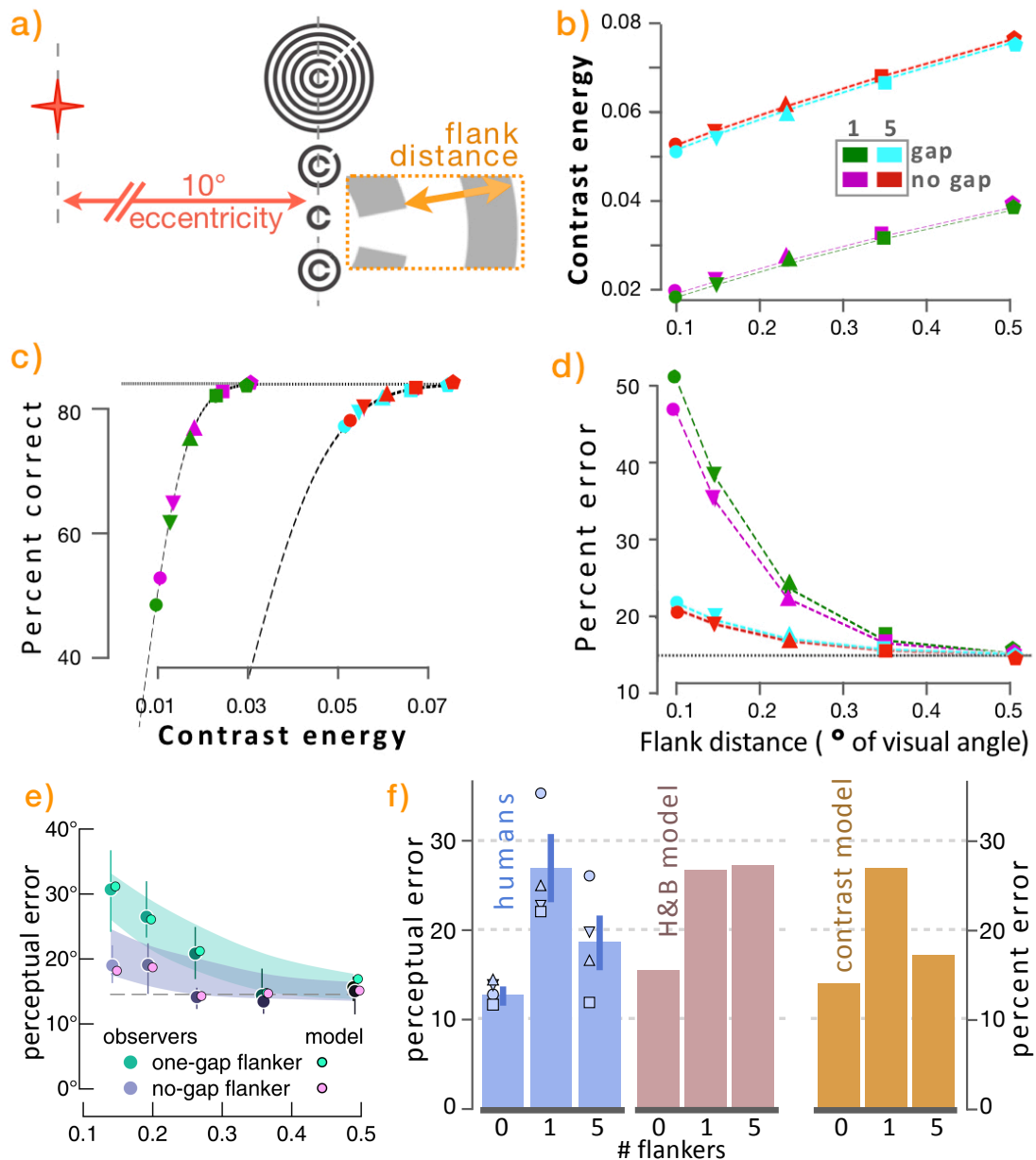

*Figure 8.  **a)** Sample stimuli: Landolt C target either alone or encircled by different numbers of (differently oriented) larger Landolt Cs (or by un-gapped full circles).  Humans are asked to manually rotate a prompting image to correspond to the orientation of the gap in the target C.  **b)** Ungapped flankers elicit higher contrast values for every flank distance and for every number of flankers (red above aqua; purple above green).  Contrast energy is substantially larger for images with five flankers (red, aqua) than with one flanker (purple, green).  (Note, when gap size is increased, the difference between gap and no-gap conditions also increases (not shown).  **c)** Sensitivity to contrast differences is steeper for images with a single flanker than for those with five flankers; images with five flankers are almost all conducive to target object recognition.  **d)** Combining b) and c), the most robust target recognition occurs for images with five flankers, and gapped flankers induce errors modestly but reliably more than flankers with no gaps.  **e)** Corresponding gap results from the original studies (from H&B 2015 Fig 1d), showing that humans make more errors when flankers have gaps.  **f)** Effect of different numbers of flankers on human recognition as reported in Pachai et al. 2016 (light blue), on the model introduced by Harrison and Bex 2015 (red), and on the contrast energy calculation introduced here (orange).*

The contrast predictions (Figure 8d) somewhat predict the direction and shape of the reported empirical findings from the published article (reproduced in Figure 8e), although the quantities clearly differ.

Figure 8e also shows a good fit to the empirical findings of a model introduced by the authors. That model is further tested in panel 8f, which shows the effects of different numbers of flankers on human and model recognition. We reproduce their results together with the predictions made by the contrast energy metric.

Human recognition (light blue) empirically exhibits lower error (less crowding) for 0 and for 5 flankers, than for a single (1) flanker; the bars show mean and s.d., and the individual results of the four subjects are shown as individual small symbols. Pachai et al. (2016) showed that the model introduced by Harrison and Bex (2015) erroneously predicts that crowding will be roughly the same whether there is a single flanker or five flankers (red bars).

The contrast metric generates predictions that are more in line with empirical findings: five flankers elicit less crowding than a single flanker (orange). This is simply because the added flankers substantially increase the generalized contrast of the overall image to be viewed by the subject, and higher contrast predicts better recognition.

Once again, it is not clear whether this seemingly too-simple explanation is accurately reflecting the vision mechanisms of the viewing subject, but it clearly shows that contrast alone does match these particular aspects of the empirical findings. Contrast is an explanation at best, a confound of the experiment at worst.

It is worth noting that Harrison and Bex (2015) subsequently modified their model (Harrison and Bex 2016) to conform to the findings pointed out by Pachai et al. (2016), perhaps further illustrating the difficulty of simplifying or distilling such models into simpler explanatory principles.

We emphasize the unexpected nature of these findings. The contrast energy model presented here intentionally omits any information about object shapes, angles, gaps. This of course should be expected to prevent the model from predicting the experimental findings, since the features that are specifically being measured in those experiments (such as Landolt C gap angle) are being entirely ignored in the contrast energy model.

Yet what we find is that the simple contrast energy model produces predicted response accuracies that appear to qualitatively correspond with subjects' response accuracies, even though the model is clearly not, and cannot be, carrying out the task that the subjects are purportedly accomplishing. Rather, contrast energy alone predicts reduced response accuracies in a way that tracks subjects' response accuracies, suggesting that either the human experimental results may actually be due to something other than they are intended to, or at least that they may be confounded by these contrast measures.

As seen in Figure 8f, response accuracies are reduced (errors increased) by the addition of one flanker to the Landolt C, but those reduced accuracies are ameliorated by the addition instead of five (rather than one) flankers. The leftmost graph shows this for human subjects (Pachai et al., 2016). The next graph shows that Pachai et al.'s use of the model of Harrison and Bex (2015) does not predict this non-monotonic effect; the H&B model predicts that five flankers will yield roughly as many errors as one flanker. Both of those graphs exhibit "perceptual error" on the y axis in the form of the SD of a Von Mises function fit to the distribution of errors across trials, corresponding to subjects' reports of the orientation of the target Landolt C gap (Harrison & Bex 2015). The rightmost graph, however, exhibits no reference to the C gap orientation, but rather reports the percent error that is predicted solely from the calculation of contrast energy values in the visual materials of each of the experiments. Yet these measures appear to qualitatively correspond to the human measures: five flankers produce less crowding interference than one flanker.

Many features may cause crowding, but the present model proceeds from the prediction that contrast will itself affect response accuracy, independent of other features. We are forwarding the possibility that this is a confound that may affect the interpretation of the identified results. That is, it is not yet ruled out that the magnitude of the angle error differences may be due wholly or in part to contrast energy differences, which change with the flanker features in these specific experiments.

## F) Two summary tables

| exper | ecc | dist (cm) | pixels/degree | contrast_sigma | mu_t | sigma_t | E_alpha | k_phi |
|-------|-------|-----------|---------------|----------------|----------|-----------|---------|---------|
| HE    | 3.88  | 75.00     | 46.4177       | 0.5541         | 0.020228 | 0.0006665 | 0.080   | 29.5499 |
| KA    | 5.00  | 2300.00   | 1423.4770     | 10.8793        | 0.016057 | 0.0005291 | 0.032   | 9.9290  |
| PA1   | 10.00 | 58.00     | 35.8963       | 0.8324         | 0.011640 | 0.0003835 | 0.030   | 15.7745 |
| PA5   | 10.00 | 58.00     | "             | "              | "        | "         | 0.080   | 58.7319 |
| PL    | 5.00  | 55.88     | 34.5843       | 0.4956         | 0.012000 | 0.0003954 | 0.010   | 10.8333 |
| PL    | 10.00 | 55.88     | "             | 0.7620         | "        | "         | 0.010   | "       |
| PL    | 15.00 | 55.88     | "             | 1.0688         | "        | "         | 0.010   | "       |
| PL    | 20.00 | 55.88     | "             | 1.4252         | "        | "         | 0.010   | "       |
| PS    | 5.00  | 55.88     | "             | 0.4956         | 0.005000 | 0.0001648 | 0.025   | 9.9999  |
| PS    | 10.00 | 55.88     | "             | 0.7620         | "        | "         | 0.025   | "       |
| PS    | 15.00 | 55.88     | "             | 1.0688         | "        | "         | 0.025   | "       |
| PS    | 20.00 | 55.88     | "             | 1.4252         | "        | "         | 0.025   | "       |

Table 1. Numerical values for all reported experiments.

Table 1 lists the numerical constants used in each of the experimental analyses. The eccentricities and distances are given in the experiments. Pixels per degree is determined by measuring monitor resolution (which was 0.282 mm/pixel on the monitor used in all experimental analyses reported here). Contrast sigma values come directly from Dacey & Peterson's measures of midget cell dendritic radii (Dacey and Petersen 1992). Mu_t ( $\mu_\tau$ ) is the measured contrast energy of the target alone in any given experiment. Sigma_t ( $\sigma_\tau$ ), as described earlier, is calculated such that 10% variation from mu_t will yield a predicted 1% identification rate of the target. E_alpha ( $E_\alpha$ ), as described, is the contrast energy value at which the target is identified 99% percent of the time by a subject in a given experiment; this value is the sole parameter that is measured from subjects' behaviors in the experiment to be analyzed. K_phi is the ratio $k_\phi = \sigma_\phi/\sigma_\tau$ as described; it is thus a function of the value of $E_\alpha$. It is notable that, for instance, the distance from subject to image is constant in many of the experiments, varies a little in Pachai and Manassi, and then is orders of magnitude different in the Flom et al. experiments. The pixels per degree thus vary by an equally large amount, and the derived $k_\phi$ ratio thus varies somewhat. All of these numerical values are derived directly from the experiments themselves, other than $E_\alpha$, as described earlier. That parameter alone is shown to be capable of capturing these order-of-magnitude differences in experimental design across the experiments analyzed here, suggesting the potentially broad generality of the approach that is forwarded here. (Experiments listed: HE (Manassi et al., 2012); KA (Flom et al., 1963); PA1,5 (Harrison & Bex 2015; Pachai et al., 2016); PL, PS (Pelli & Tillman 2008)).

|  | confounded with RMS image contrast | crowding predicted by radially generalized contrast |
|---|---|---|
| Pelli, Tillman '08 fig 5 | no | yes |
| Flom et al '63 exp.1 | no | yes |
| Manassi et al. '12 exp.1 | • number of flanks increases contrast | Fig 1a,b,c; not Fig 1d |
| Pachai et al. '16 exp.1 | • number of flanks increases contrast<br>• flank distance increases contrast | yes |

Table 2. Instances of confounds and predictions in the analyzed experiments.

In all the experiments analyzed, there are reported crowding effects that are predictable from the radially generalized contrast metric that has been introduced (Table 2). Manassi et al. '12 provide instances of experiments some of which yield equivocal results in our analyses. Of particular note is their Figure 1d, in which flankers occur at staggered locations; the contrast energy quantity predicts the same outcome for this as for their Figure 1a. The jittered arrangement of flankers in that image has an effect on the subject that is not due solely to contrast. (In fact, as mentioned, the flankers now resemble verniers, possibly causing subjects to fail to identify the *location* of the target.)

Notably, in both Manassi et al. and Pachai et al., there are reported crowding effects that increase directly even with the standard RMS contrast of the image. As flankers are added, pixels are added, and the overall contrast energy of the image increases. (In Pachai et al., the posited reason that flank distance increases contrast is because farther flanks become larger circles, which require more pixels, unlike many other forms of flankers.)

The present work studied changes to the overall contrast and spatial frequency characteristics of the *entire* image being viewed by the subject, showing that such changes directly affect crowding. Many other factors remain: for instance, the dissimilarity of target and flankers can affect crowding, mediated at least in part by contrast, e.g., (Kooi, Toet et al. 1994), and in some cases the spatial frequency of just the target, or just the flankers, themselves, may be changed without changing the amount of crowding (see, e.g., (Chung, Levi et al. 2001)). These and many additional factors may interact with the total generalized contrast of a field of view, presenting opportunities for further studies on multiple fronts.

## IV. Discussion
### A) Contrast is a principal factor in some forms of visual crowding
A novel generalized extension of visual contrast energy, the calculated semi-norm $C(\bar{v})$, (Eq.4) generated values that correspond to the basic findings in a range of reported experimental results on visual crowding. This surprisingly suggests that contrast alone substantially confounds many of these published results, i.e., contrast provides an explanatory account of some key characteristics of crowding.

There exist multiple further effects deemed crowding that likely are not simply contrast-based. We showed one such example from Manassi et al (2012); there are many more in the literature (see, e.g., (Manassi and Whitney 2018)). The results presented here may thus suggest that the field divide findings between those accounted for by the contrast energy semi-norm presented here versus those that are not. It is possible these two categories of psychophysical effects should be treated separately, rather than unified: contrast-dependent crowding vs. contrast-independent crowding.

To date, accounts of crowding have grown to incorporate extensive simulations of multiple interacting neural elements in hierarchical architectures (Freeman and Simoncelli 2011, Harrison and Bex 2015). These complex systems typically reproduce the phenomena of crowding with some accuracy, though sometimes requiring modification when challenged with novel findings (Harrison and Bex 2015, Herzog, Sayim et al. 2015). Some of these approaches in the literature have prominently been based on successive "pooling," in which early units with relatively tight receptive fields project to those with larger RFs, combining (pooling) the information and thus disarranging features in the stimulus (Freeman and Simoncelli 2011), concordant with computational models that entail successive hierarchical stages of visual processing (Riesenhuber and Poggio 1999, Rodriguez, Whitson et al. 2004). Other approaches are based on limitations of a hypothesized attentional resource, e.g., (He, Cavanagh et al. 1996, Intriligator and Cavanagh 2001), preventing adequate perception of crowded stimuli.

The crowding effect typically arises from the addition of flankers to target images; however, surprisingly, there are cases where adding flankers can paradoxically improve target identification (Banks, Larson et al. 1979, Wolford and Chambers 1983, Manassi, Sayim et al. 2012, Herzog, Sayim et al. 2015), and that flankers that are far outside of what is usually considered the effective neighborhood (Bouma 1970) of the target nonetheless can affect crowding (Manassi, Sayim et al. 2012, Manassi, Sayim et al. 2013).

It has been proposed (Herzog, Sayim et al. 2015) that these seemingly paradoxical effects (of multiple flankers and of distant flankers) may be due to "grouping" of flankers together, with these "groupings" somehow causing the multiple flankers to affect each other more than affecting the target. Such an effect appears to suggest that the results arise from downstream "configural" processing, beyond early visual regions (Chicherov, Plomp et al. 2014). This is an intriguing topic of ongoing study.

The findings in the present report indicate that simulations of multiple interacting neural-like elements may in large part be replaced by a relatively straightforward formulation (Eqns 4, 5) to calculate visual contrast and apply it to behavioral recognition. We conjecture that some of the heretofore hidden fundamental principles underlying some complex simulation findings are being captured by the generalized visual contrast metric introduced here.

All of the code used throughout this article is available on github, to encourage further experimentation with the work that is presented here.
    https://github.com/DartmouthGrangerLab/Contrast/

There may likely be additional downstream visual computations beyond those of contrast alone; it will be of interest to see whether these constitute a separate natural category of crowding phenomena. These and several other classes of experiments are being pursued as further investigations of relations between crowding and contrast.

### B) Sources of contrast information in the visual stream

The generalized contrast calculation introduced here is calculated from measures of dendritic radii in midget cells in the retina; corresponding center-surround phenomena also occur in the thalamus and cortex; further experiments will pursue the question of which of the many Gaussian and/or center-surround operations along visual pathways may participate in these metric calculations.

There are other formulations of contrast that may play a similar role. Contrast response functions from the literature combine a measure of psychophysical contrast plus a behavioral mapping function; one such measure of center-surround interactions calculated the contrast values of a target and a surround ($C_s$, $C_t$), and then fitted a response model

$$R_t = \frac{k(1+W_e C_s^{pe})C_t^p}{(1+aC_t^q+W_i C_s^{qi})}$$

by estimating seven parameters: $p, q, a, W_e, W_i, pe, qi$ (Xing and Heeger 2001). Such models may, after suitable fitting, also account for crowding data of the kind addressed here. It will be of interest to pursue these possibilities in further studies.

The examples in the present paper arise from simple center-surround interactions at apparently relatively early stages of the visual stream. Higher-level "configural" effects that also appear in the crowding literature may arise from completely different sources than contrast, or possibly could arise from combinations of successive center-surround operations, compositing these into more complex organizations. It may prove possible to distinguish early contrast-dependent effects from other more downstream crowding effects; if so, this perhaps may enable the nomenclature of crowding to be revised to reflect such a distinction.

Recognizing is not simply perceiving; it is, further, matching the perceived entity (a letter, a target C, etc.) against some specific stored memory or template from the experiment's instructions. Subjects still "perceive" the existence of, say, a crowded "r", but they fail to identify what letter those perceived pixels connote; i.e., to match the perceived image against some predetermined knowledge of an "r" versus an "n" or "h". Small receptive fields enable foveal pixels to be processed within minute regions, retaining the relative positions of different parts of an image (e.g., an "r"), and maintaining separate processing of the target versus neighboring pixels from flankers.

As receptive field size increases with eccentricity, acuity is reduced. That peripheral acuity is still quite sufficient to recognize a letter in isolation. What fails is that the subject fails to <u>recognize</u> the letter, even though its pixels are perceived. In peripheral larger RFs, pixels within a given RF are more likely to be processed as part of a single entity, rather than separately as they would be in smaller RFs. The boundaries between pixels of the target versus flanker are lessened; thus the target is still perceived, but its identity may be obscured by interactions among pixels within too-large RFs. (See, e.g., Figures 5b and 6b). The radially-generalized contrast energy introduced here, essentially predicts simply that recognition is assisted by large contrast differences.

(This predicts that crowding should continue to occur for any RF size, even foveal, as long as the closeness of the flankers is scaled according to RF size. It is worth noting that evidence for this has been provided in the literature; see, e.g., Coates et al., 2018).

If early crowding were actively *preventing* recognition of target objects, then information would be lost, and presumably could not be available for further (downstream) processing; yet some experiments appear to show just such downstream availability (see, e.g., (Manassi and Whitney 2018)). The results presented here suggest that information about an object is not "lost"; it simply is one type of information (contrast) that can become input to a (downstream) behavioral mapping process; when this data is queried, it may be insufficient to answer certain questions, yet may still provide an otherwise unimpeded stream of feature information that is available to other decision queries that may occur. These operations too are of interest as topics of further study.

Of related interest is recent work showing that standard feedforward convolutional neural networks (ffcnns) are in principle incapable of producing global shape computations that are shown to be used in human visual processing; some alternative models to ffcnns are in extended development that avoid the ffcnn pitfalls (Doerig, Bornet et al. 2019, Doerig, Bornet et al. 2020).

### *C) Implications of the transform from physical to perceptual space*
Visual contrast is at the root of recognition (e.g., (Strasburger and Rentschler 1996, Avidan, Harel et al. 2002)). We have here proposed that standard measures of contrast can be generalized to incorporate spatial relations among pixels, such that influences of neighbors are included, as in Figure 2a, b, and e. The resulting Radially-Generalized Contrast is proffered as a fundamental explanatory metric that predictively addresses spatial relations in visual processing. We have shown that standard quantities such as RMS contrast fall out as special cases of this novel contrast energy formulation.

We find that several well-studied instances of impeded recognition attributed to crowding arise directly from this measure of contrast energy. Either i) some reported crowding effects are critically confounded by surprisingly simple changes in contrast in the stimuli, or ii) some forms of crowding are, in essence, predominantly a contrast effect. Again, there may also be a downstream additional crowding effect that is not so directly linked to contrast. In this paper, we show that at least some of the psychophysical phenomena that have in the past been attributed to downstream effects may, unexpectedly, be largely accounted for by contrast effects alone.

The broadened contrast energy metric arises from an approach that situates image perception in an "image space" that has different distance metrics from those of the physical image itself (Bowen, Rodriguez et al. 2020). In particular, standard physical measures (and some proposed perceptual measures such as RMS contrast) typically treat separate pixels as independent (corresponding to separate dimensions in Euclidean vector representations); whereas the perceptual metric proposed here formally incorporates information about neighboring regions or pixels. These perceptual neighbor effects have been shown to have implications in image similarity, as well as their influence in crowding. Further work shows that these metrics are also in evidence in evaluations of auditory stimuli (Oh, Bowen et al. 2020), indicating that these principles may generalize even across modalities. Ongoing studies are pursuing these and additional consequences of these new contrast formulations, in a range of perceptual and cognitive paradigms.

*Acknowledgments*: This work was supported in part by grants N00014-16-1-2359 and N00014-19-1-2434 from the Office of Naval Research. Sincere gratitude for assistance with formal notation from Damian Sowinski and Eli Bowen, and thanks to Eli Bowen, Eva Childers, and Annie Brown, for advice and for contributions to ongoing experimental work on visual psychophysics.

## *References*

Avidan, G., M. Harel, T. Hendler, D. Ben-bashat, E. Zohary and R. Malach (2002). "Contrast sensitivity in human visual areas and its relationship to object recognition." J Neurophysiology **87**: 3102-3116.

Banks, W., D. Larson and W. Prinzmetal (1979). "Asymmetry of visual interference." Perception & Psychophysics **25**: 447-456.

Banks, W. and H. White (1984). "Lateral interference and perceptual grouping in visual detection." Perception & Psychophysics **36**: 285-295.

Bouma, H. (1970). "Interaction effects in parafoveal letter recognition." Nature **226**: 177-178.

Bowen, E., A. Rodriguez, D. Sowinski and R. Granger (2020). "Visual stream connectivity predicts assessments of image quality." arxiv.org/abs/2008.06939

Chicherov, V., G. Plomp and M. Herzog (2014). "Neural correlates of visual crowding." NeuroImage **93**: 23-31.

Chung, S., D. Levi and G. Legge (2001). "Spatial-frequency and contrast properties of crowding." Vision Research **41**: 1833-1850.

Coates, D., D. Levi, P. Touch and R. Sabesan (2018). "Foveal crowding resolved." Scientific Reports **8**: 9177.

Dacey, D. (2004). Origins of perception: retinal ganglion cell diversity and the creation of parallel visual pathways. The Cognitive Neurosciences.

Dacey, D. and M. Petersen (1992). "Dendritic field size and morphology of midget and parasol ganglion cells of the human retina." Proceedings of the National Academy of Science **89**: 9666-9670.

Doerig, A., A. Bornet, O. Choung and M. Herzog (2020). "Crowding reveals fundamental differences in local vs. global processing in humans and machines. ." Vision Research **167**: 39-45.

Doerig, A., A. Bornet, R. Rosenholtz, G. Francis, A. Clarke and M. Herzog (2019). "Beyond Bouma's window: How to explain global aspects of crowding?" PLoS Comput Biol **15**: e1006580.

Flom, M., F. Weymouth and D. Kahneman (1963). "Visual resolution and contour interaction." Journal of the Optical Society of America **53**: 1026

Freeman, J. and E. Simoncelli (2011). "Metamers of the ventral stream." Nat Neurosci **14**: 1195-1201

Harrison, W. and P. Bex (2015). "A unifying model of orientation crowding in peripheral vision." Current Biology **25**: 3213-3219.

Harrison, W. and P. Bex (2016). "Reply to Pachai et al." Current Biology **26**: R353-R354.

He, S., P. Cavanagh and J. Intriligator (1996). "Attentional resolution and the locus of visual awareness " Nature **383**: 334-337.

Herzog, M., B. Sayim, V. Chcherov and M. Manassi (2015). "Crowding, grouping, and object recognition." Journal of Vision **15**: 1-18.

Intriligator, J. and P. Cavanagh (2001). "The spatial resolution of visual attention." Cognitive Psychology **43**: 171-216.

Kooi, F., A. Toet, S. Tripathy and D. Levi (1994). "The effect of similarity and duration on spatial interaction in peripheral vision." Spatial Vision **8**: 255-279.

Kukkonen, H., J. Rovamo, K. Tiippana and R. Nasanen (1993). "Michelson contrast, RMS contrast and energy of various spatial stimuli at threshold." Vision Research **33**: 1431-1436.

Legge, G. (2007). The psychophysics of reading in normal and low vision, Erlbaum.

Levi, D., S. Hariharan and S. Klein (2002). "Suppressive and facilitatory spatial interactions in peripheral vision: Peripheral crowding is neither size invariant nor simple contrast masking. ." Journal of Vision **2**: 167-177.

Manassi, M., B. Sayim and M. Herzog (2012). "Grouping, pooling, and when bigger is better in visual crowding." Journal of Vision **12**: 1-14.

Manassi, M., B. Sayim and M. Herzog (2013). "When crowding of crowding leads to uncrowding." Journal of Vision **13**: 1-10.

Manassi, M. and D. Whitney (2018). "Multi-level crowding and the paradox of object recognition in clutter." Current Biology **28**: R127-R133.

Oh, S., E. Bowen, A. Rodriguez, D. Sowinski, E. Childers, A. Brown, L. Ray and R. Granger (2020). "Towards a perceptual distance metric for auditory stimuli." https://arxiv.org/abs/2011.00088

Pachai, M., A. Doerig and M. Herzog (2016). "How best to unify crowding?" Current Biology **26**: R343-R345.

Parkes, L., J. Lund, A. Angeluicci, J. Solomon and M. Morgan (2001). "Compulsory averaging of crowded orentation signals in human vision." Nat Neurosci **4**: 739-744.

Peli, E. (1990). "Contrast in complex images. ." J Optical Society of America **7**: 2032-2040.

Pelli, D. (2008). "Crowding: A cortical constraint on object recognition." Curr Opin Neurobiol **18**: 445-451.

Pelli, D., M. Palomares and N. Majaj (2004). "Crowding is unlike ordinary masking: distinguishing feature integration from detection." Journal of Vision **4**: 1136-1169.

Pelli, D. and K. Tillman (2008). "The uncrowded window of object recognition." Nat Neurosci **11**: 1129-1135.

Riesenhuber, M. and T. Poggio (1999). "Hierarchical models of object recognition in cortex." Nat Neurosci **2**: 1019-1025.

Rodieck, R. (1965). "Quantitative analysis of cat retinal ganglion cell response to visual stimuli." Vision Research **5**: 583-601.

Rodriguez, A., J. Whitson and R. Granger (2004). "Derivation and analysis of basic computational operations of thalamocortical circuits." Journal of Cognitive Neurosci **16**: 856-877.

Strasburger, H. and I. Rentschler (1996). "Contrast-dependent dissociation of visual recognition and detection fields." Eur J Neurosci **8**: 1787-1791.

Strasburger, H., I. Rentschler and M. Jüttner (2011). "Peripheral vision and pattern recognition: a review." Journal of Vision **11**: 1-82.

Wandell, B. (1995). Foundations of Vision, Sinauer.

Watson, A., H. Barlow and J. Robson (1983). "What does the eye see best? ." Nature **302**: 419-422.

Wolford, G. and L. Chambers (1983). "Lateral masking as a function of spacing." Perception & Psychophysics **33**: 129-138.

Xing, J. and D. Heeger (2001). "Measurement and modeling of center-surround suppression and enhancement." Vision Research **41**: 571-583.

Young, R. (1987). "The Gaussian derivative model for spatial vision: I. Retinal mechanisms." Spatial Vision **2**: 273-293.